\documentclass[twocolumn,preprintnumbers,amsmath,amssymb]{revtex4}

\usepackage{graphicx}
\usepackage{bm}
\usepackage{dcolumn}
\usepackage{color}

\newcommand{\be}{\begin{equation}}
\newcommand{\ee}{\end{equation}}
\newcommand{\bea}{\begin{eqnarray}}
\newcommand{\eea}{\end{eqnarray}}
\newcommand{\non}{\nonumber}

\begin{document}

\title{On infrared and ultraviolet divergences of cosmological perturbations}

\author{Giovanni Marozzi}
\email{giovanni.marozzi@college-de-france.fr}
\affiliation{Coll\`ege de France, \\ 11 Place M. Berthelot,  75005 Paris, France}
\author{Massimiliano Rinaldi}
\email{massimiliano.rinaldi@unige.ch}
\affiliation{D\'epartment de Physique Th\'eorique,  
Universit\'e de Gen\`eve, \\24 quai E.\ Ansermet  CH--1211, Gen\`eve 4, Switzerland}
\author{Ruth Durrer}
\email{ruth.durrer@unige.ch}
\affiliation{D\'epartment de Physique Th\'eorique and Center for Astroparticle Physics,  
Universit\'e de Gen\`eve, \\24 quai E.\ Ansermet  CH--1211, Gen\`eve 4, Switzerland\\ and \\ CEA, SPhT, URA 2306, F-91191 Gif-sur-Yvette, France
}

\begin{abstract}

\noindent We study a consistent infrared and ultraviolet regularization scheme for the cosmological perturbations. The infrared divergences are cured by assuming that the Universe undergoes a transition between a non-singular pre-inflationary, radiation-dominated  phase and a slow-roll inflationary evolution. The ultraviolet divergences are eliminated via adiabatic subtraction. 
A consistent regularization of the field fluctuations through this transition is obtained by
performing a mode matching for both the gauge invariant Mukhanov variable and its adiabatic expansion. We show that these quantities do not generate ultraviolet divergences other than the standard ones, when evolving through the matching time. 
We also show how the de Witt-Schwinger expansion, which can be used to construct the counter-terms regularizing the ultraviolet divergences, ceases to be valid well before horizon exit of the scales of interest. 
Thus, such counter-terms should not be used beyond the time of the horizon exit and
it is unlikely that the observed power spectrum is modified by adiabatic subtraction, as claimed in some literature. On the contrary, the infrared regularization might have an impact on the observed spectrum, and we briefly discuss this possibility.

\end{abstract}


\maketitle


\section{Introduction}


\noindent According to the inflationary paradigm, the large scale structure observed today in the sky  originated from tiny quantum fluctuation 
in the early Universe. At that time, the Universe experienced a period of quasi-exponential expansion that amplified these fluctuations, which became classical after exiting the horizon  and generated the gravitational instabilities responsible for the formation of  the large-scale structures. One of the most important predictions of the inflationary scenario is that the spectrum of these fluctuations is nearly scale-invariant \cite{infl}. 

The quantum origin of the perturbations of the inflaton field and of the  metric tensor necessarily raises a concern about renormalization. In fact, it is well known that the correlation functions of quantum fields on a curved background suffer from divergences that cannot be cured as in flat space \cite{BD}. In the case of a time-dependent Friedmann-Lemaitre-Robertson-Walker (FLRW) background, quantum correlations typically develop logarithmic divergences  in the ultraviolet (UV), together with the familiar quadratic ones. In addition, also infrared (IR) divergences exist. These problems become quite relevant, as the  
observed power spectrum is directly connected to the two-point function of the gauge invariant Mukhanov variable \cite{Mukhanov} in the coincidence limit.

Concerning the UV divergences, the infinities can be cancelled by subtracting counter-terms constructed according to the adiabatic expansion in the momentum space \cite{ParkerAE}, which is equivalent to the de Witt-Schwinger technique in coordinate space \cite{BunchParker}. According 
to~\cite{ParkerFirst,Parker2, Agullo},  the adiabatic subtraction leaves an imprint on
the renormalized power spectrum because the adiabatic counter-terms, when evaluated a few Hubble time after the horizon exit, are relatively large for the scales of interest.
On the other hand, in a recent paper \cite{us}, we argued that this is not correct. Among other problems, the main one is that the adiabatic subtraction procedure seems to be ill-defined when the scales of interest are stretched towards horizon exit  (see also \cite{FMVVPRD76} for a different criticism on the main idea proposed in \cite{ParkerFirst}). At the end of this paper, we offer a further explanation on why the adiabatic subtraction should not be considered valid around horizon crossing. 

Regarding the cure for IR divergences one possibility consists in assuming an initial vacuum state which differs from the usual Bunch-Davies vacuum. Note that this approach is different from what was done in the context of the trans-Planckian problem, see e.\ g.\   \cite{alfavac}, where the interest was about the UV behavior only. Physically, this is equivalent to assume that the Universe emerges from a pre-inflationary phase dominated, for example, 
by matter or radiation, see e.\ g.\  \cite{proko,proko2} \footnote{Note that the
assumption that inflation is not eternal in the past has also been used in
\cite{FMVV2002} to justify the appearance of a natural infrared cut-off in
the problem.}. 
A preinflationary matter-dominated phase, and its
effect on the CMB, was also considered in \cite{SGC2011}.
Then, in line with an old theorem formulated by Ford and Parker \cite{ford}, no IR divergences can develop during the subsequent expansion. In this paper, we assume that fluctuation modes evolve across a sharp transition from a radiation-dominated Universe to an inflationary  slow-roll phase \footnote{A 
similar scheme was used in
\cite{Vilenkin:1982wt} for the infrared problem associated to a massless
minimally coupled scalar field in a de Sitter space-time.}. 
The only requirement that we impose is the continuity of the scale factor and of its first derivative 
as in \cite{proko}. 
Mode matching has been already utilized to  study observational signatures of pre-inflationary phases characterized by lower-dimensional effective gravity or modified dispersion relation in~\cite{maxnew}.
 Also, the spectrum of gravitational waves generated by a series of  radiation and matter dominated phases has been studied~in \cite{maxruth}.

The main question that we want to address here is whether the UV and IR renormalization schemes outlined above can be performed together. Specifically, first we show that the new terms that arise from a non-trivial vacuum, and that regularize the IR divergences, do not generate, after the match, new UV divergences with respect to the standard ones which characterize the inflationary phase. 
Then, we want to verify if also the adiabatic counter-terms, that must be present already before the match, evolve consistently through the match.

The plan of the paper is the following. In Sec.\ \ref{due} we introduce the formalism.
In Sec.\ \ref{tre} we outline the mode matching technique applied to the scalar perturbations in terms of the Mukhanov variable. 
In Sec.\ \ref{quattro} we study the propagation of the IR-regulating terms through the phase transition at the matching time. In Sec.\ \ref{cinque} we extend this analysis to the adiabatic counter-terms. In Sec.\ \ref{sei} we estimate the potential impact of these new terms in the prediction of the inflationary spectra. As mentioned above, in Sec.\ \ref{sette} we show why adiabatic subtraction should not be taken too seriously when the scales of interest to us cross the horizon. We finally conclude in Sec.\ \ref{otto} with a summary of our results.


\section{Formalism}\label{due}


\noindent Let us consider a spatially flat Universe, whose dynamics is driven by 
a classical minimally coupled scalar field, and is described by the action 
    \be
    S = \int d^4x \sqrt{-g} \left[ 
\frac{R}{16{\pi}G}
    - \frac{1}{2} g^{\mu \nu}
    \partial_{\mu} \phi \partial_{\nu} \phi
    - V(\phi) \right]\ .
    \label{action}
\ee
For a spatially flat FLRW spacetime, with metric
\bea\label{metric}
ds^2=-dt^2+a^2(t)\delta_{ij}dx^idx^j\ ,
\eea 
the background equations of motion 
for $\phi(t)$ and for the scale factor $a(t)$ read
\begin{eqnarray}
\ddot{\phi}+3H\dot{\phi}+V_\phi&=&0 \label{eqmotion1}\ ,  \\
\left(\frac{\dot a}{a}\right)^2=H^2 = {1 \over 3 M_{\rm pl}}\rho 
&=&\frac{1}{3 M_{\rm pl}^2} 
        \left(\frac{\dot \phi^2}{2} + V \right)\ , \label{Fried}\\
\dot{H}=-\frac{1}{2 M _{\rm pl}^2}(\rho+p)&=& -\frac{1}{2 M _{\rm pl}^2} \dot{\phi}^2\ ,
\label{EE00}
\end{eqnarray}
where $M_{\rm pl}^2=1/(8\pi G)$ is the reduced Planck mass, and 
the energy density $\rho$ and the pressure $p$ are related 
by the equation of state $p=\omega\rho$.
The dot denotes a derivative with respect to the cosmic time $t$.

The perturbations of the background metric can be written in the form
\bea
ds^{2}=-(1+2\Psi)dt^{2}+a^{2}\left[ (1-2\Phi)\delta_{ij}+h_{ij}\right]dx^{i}dx^{j}\ ,
\eea
and for single-scalar field inflationary models the Bardeen potentials $\Psi$ and $\Phi$ coincide. Analogously, also the inflaton field is perturbed according to $\phi\rightarrow\phi+\delta\phi$. One can conveniently describe the scalar perturbations by means of  the so-called Mukhanov variable $Q$, defined as  \cite{Mukhanov}
\bea
Q &=& \delta\phi+{\dot\phi\over H}\Psi  \\
   &=& \Psi + \frac{2H^{-1}\dot\Psi + \Psi}{3(1+\omega)}\ .
\eea
For the second expression, which is also well defined in a Universe dominated by a fluid, we
have used the Einstein constraint equation, see e.g.~\cite{mybook}.
Upon quantization,  one promotes the variable $Q$ to the operator
\be
    \hat{Q} (t, {\bf x}) = \frac{1}{(2 \pi)^{3/2}}\!\! \int \!\!d^3{\bf k}
    \left[ Q_{k} (t) \, e^{i {\bf k} \cdot {\bf x}} \, \hat{b}_{\bf
    k} + Q^*_{k} (t) e^{- i {\bf k} \cdot {\bf x}}
    \hat{b}^\dagger_{{\bf k}} \right],
    \label{quantumFourier}
    \ee
where $\hat{b}_k$ is a time-independent Heisenberg operator that satisfies the usual commutation relations
 \be
    [\hat{b}_{\bf k}, \hat{b}_{{\bf k}'}] =
    [\hat{b}^\dagger_{\bf k}, \hat{b}^\dagger_{{\bf k}'}] = 0\ , \quad
    [\hat{b}_{\bf k}, \hat{b}^{\dagger}_{{\bf k}'}] = \delta^{(3)}
    ({\bf k} - {\bf k}')\ ,
\ee
provided the modes $Q_k$ satisfy the Wronskian condition
\be
    Q_{\bf k} \, \dot{Q}^*_{\bf k} -
    \dot{Q}_{\bf k} Q^*_{\bf k} = \frac{i}{a^3} \ .
    \label{wronsk}
\ee
The equation of motion is then given by 
\be
\ddot{Q}_k + 3 H \dot{Q}_k + \frac{k^2}{a^2} Q_k +
\left[ V_{\phi \phi} + 2 \frac{d}{dt}\left(3 H + 
\frac{\dot H}{H}\right)\right] Q_k = 0,
\label{Eq_mukhanov}
\ee
where $V_{\phi\phi}$ denotes the second derivative of $V$ with respect to $\phi$.
We now introduce the slow roll parameters:
\be 
\epsilon=\frac{\dot\phi^2}{2M_{\rm pl}^2H^{2}} = -{\dot H\over H^2}\ , \qquad
\eta=M_{\rm pl}^2 \frac{V_{\phi\phi}}{V}\ \,.
\label{SRParameter}
\ee
Note that our definition of $\epsilon$ is more general than the usual one, 
$\epsilon=(M_{\rm pl}^2/2)(V_\phi/V)^2 $, 
as it is well-defined for arbitrary backgrounds, not only the ones dominated by a scalar field.
With these, Eq.\ (\ref{Eq_mukhanov}) can be rewritten as
\begin{eqnarray}\non
&&\ddot{Q}_k + 3 H \dot{Q}_k +
 \frac{k^2}{a^2} Q_k 
+\\&&
+H^{2}\Bigg[ 3 \eta -6 \epsilon  +2\epsilon^2
-\eta\epsilon -2
\frac{\dot{\epsilon}}{H}\Bigg] Q_k = 0 \ .
\label{Eq_mukhanov_2}
\end{eqnarray}

We note that the derivative $\dot\epsilon$ is related to $\epsilon$ and $\eta$ by
\be\label{doteps}
\dot\epsilon = -2\epsilon(\eta -2\epsilon)H \,.
\ee
This equation can be used to replace $\dot{\epsilon}$ in Eq.~(\ref{Eq_mukhanov_2}) but it
can also be used as a definition of $\eta$ for an arbitrary matter content of the Universe, 
not necessarily a scalar field. 
Eq.~(\ref{Eq_mukhanov_2}) holds  only if the degree of freedom $Q$ comes from
a scalar field. If it corresponds to fluctuations in a fluid with equation of state of the form $p=\omega\rho$, the $k^2$ term must be multiplied by the factor 
$c_s^2 =\dot p/\dot\rho$.
In all other aspects, the linear perturbation equation remains identical,
see~\cite{mybook} for more details. Finally, $Q_k$ is associated to the scalar power spectrum, defined by
\be
P_{\zeta}(k)=\frac{k^3}{2 \pi^2} \left(\frac{H}{\dot{\phi}}\right)^2 |Q_k|^2={k^{3}\over4\pi^{2}M_{\rm pl}^2\epsilon}|Q_k|^2\,,
\label{spectrum}
\ee
where we used Eq.\ (\ref {SRParameter}) for the second equality.

We now consider two different cases of interest for our investigation.
The first one is a slow-rolling inflationary evolution where $\epsilon \ll 1$ and where, as one can see from Eq.\ (\ref{doteps}),  
$\dot{\epsilon}={\cal O} (\epsilon^2)$ can be neglected to the leading order. The second one is the  
case when $\epsilon$ is exactly constant so that $\eta=2 \epsilon$, again because of Eq.\ (\ref{doteps}).
The latter
describes a power law expansion, as for 
the case of a radiation or matter-dominated Universe.
In both cases
the general solution to Eq.\ (\ref{Eq_mukhanov_2}) can be written in terms of Hankel functions, namely
 \bea\label{psisoln}
 Q_k=a^{-1} \left[E(k)u(z)+F(k)u^*(z)\right]\ ,
 \eea
 where
 \bea\label{umode}
 u(z)=\sqrt{\pi z\over4k}H^{(1)}_\mu(z)
 \eea
and  where we have introduced the new variable
 \bea\label{zvar}
 z(t,k)={k\over (1-\epsilon) a H}\ .
 \eea
 The index of the Hankel function is given by (we neglect contributions which are subleading for 
 the slow-roll inflationary case or exactly zero for the case $\dot{\epsilon}=0$)
 \bea
 \mu^2=\left[3-\epsilon\over 2(1-\epsilon)\right]^2-3 \eta +6 \epsilon \, . 
 \label{mu2}
 \eea
The Wronskian condition (\ref{wronsk}) implies that $|E(k)|^2-|F(k)|^2=1$.
 
\begin{figure}[ht]
\centering
\includegraphics[width=50mm]{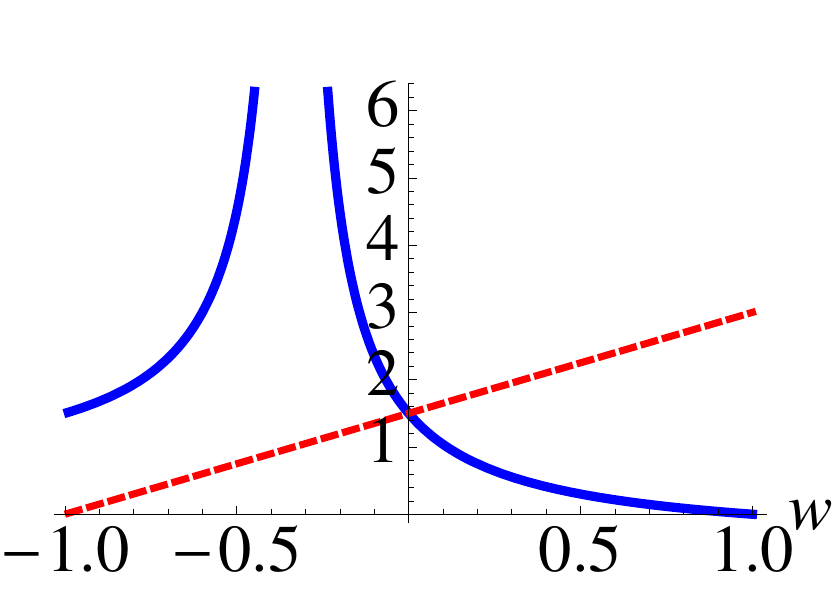}
\caption{\label{f:wepsmu}
We plot  $\epsilon$ (red dashed line) and $|\mu|$ (blue solid line) as functions of $\omega$ for power law expansion.}
\end{figure}

When $\epsilon$ is constant, it is related to the equation of state via $\epsilon={3\over 2}(1+\omega)$. In Fig.\ \ref{f:wepsmu} we plot $\epsilon$ and $\mu$ as functions of $\omega$. Note the divergence 
 at $\omega=-1/3$, which corresponds to a curvature dominated universe. In this case $\epsilon=1$ 
 and $z$ is not well defined.
Note also that, since  $\epsilon=1-{a\ddot a/ \dot a^2}$, we have that $\epsilon<1$ for an accelerating Universe and  $\epsilon>1$ for an decelerating one. Typical cases  are
 \begin{itemize}
 \item{Radiation-dominated:} $\omega=1/3$ and $\epsilon=2$,  $\mu= 1/2$. 
 \item{Matter-dominated:} $\omega=0$ and $\epsilon=3/2$, $\mu= 3/2$.
 \item{$\Lambda$-dominated:} in the slow-roll approximation we have typically, at the leading order, $\mu=3/2-\eta+3\epsilon$ with $0<\epsilon\ll 1$ and $0<|\eta|\ll 1$. 
 Thus the Universe accelerates. 
 For the exact de Sitter solution, one has $\omega=-1$, $\epsilon=\eta=0$ and $\mu=3/2$.
 \end{itemize} 
The first two cases can be obtained either by a fluid with the corresponding equation of state or by  a  scalar field with exponential potential, which yields a power-law scale factor (see, for example, \cite{Marozzi}). 

We remark that (only in four space-time dimensions) $\mu=3/2$ represents a ``degenerate'' solution, as it describes both an accelerating, de Sitter Universe, and a decelerating, matter-dominated Universe. In this case, one must specify also $\omega$ or $\epsilon$. Note also that, when the expansion of the Universe decelerates one obtains $z<0$. Therefore, the solution (\ref{umode}) should be modified to
 \bea\label{umode2}
 u(z)=\sqrt{-\frac{\pi z}{4k}}H^{(2)}_\mu(-z)\ ,
 \eea
 as one usually takes the negative axis as the branch cut for Hankel functions and one has to satisfy 
 the Wronskian condition. 
 
During inflation, the wavelength of a mode $k$ is growing, $\lambda =(2\pi/k)a(t)$, while the Hubble parameter remains nearly constant, hence $z$ is decreasing. A typical mode is inside the horizon at early times, $z\gg 1$ and crosses the horizon as inflation unfolds. For such a mode we can argue that it is initially in the Minkowski vacuum, $u(z)\propto \exp(ik\eta)/\sqrt{2k} = \exp(-iz)/\sqrt{2k} $. 
This leads to the inflationary initial condition given by $Q_k(z) = \frac{u(z)}{a}$, $E_\mu =1$ and $F_\mu =0$. 

However, during a decelerated expansion, $|z|$ is growing and
modes which are inside the horizon when inflation begins, have been outside during the decelerated expansion before inflation, and in general we have no way to determine their initial condition. For large scales, time dependence cannot be neglected and for general time dependent spacetimes we cannot formulate vacuum initial conditions. 

Fortunately there is one exception to this rule and this is the radiation dominated Universe: in this case,  we can redefine our perturbation variable to 
\be
v = aQ
\ee
such that Eq.~(\ref{Eq_mukhanov_2}), in terms of conformal time $\eta$ defined by $d\eta = dt/a(t)$,
reduces to the simple Minkowski wave equation
\be\label{Minkwe}
v'' + k^2v = 0 \ .
\ee
Here a prime denotes the derivative w.r.t. conformal time $\eta$.
(Again, if we replace the scalar field leading to an radiation-dominated expansion law, $a\propto t^{1/2}$, with a radiation fluid, we must substitute  the term $k^2$ by $k^2/3$.) As we know how to quantize Eq.\ (\ref{Minkwe}), we can set up Minkowski vacuum initial conditions for all modes. Here the expansion of the universe has disappeared and is taken into account simply by the normalization, $v=aQ$ and $dt = ad\eta$. The vacuum initial condition for $v$ corresponds exactly to
Eq.~(\ref{umode2}) for $\mu=1/2$, and $v = aQ = u(z)=\sqrt{-\frac{\pi z}{4k}}H^{(2)}_{1/2}(-z) = i \exp(-ik\eta)/\sqrt{2k}$. Apart from the de Sitter case, where time dependence can be fully removed, this is the only FLRW background which allows for well defined quantum initial 
conditions for all modes. The reason for that is that for a radiation dominated universe, the 
Ricci scalar, $R=\dot H + 2H^2$ vanishes. Therefore, any scalar field can be considered 
as conformally coupled and as such is not affected by the expansion of the Universe.

This observation is crucial for the model which we develop in the next section where we match
a previous radiation era to a subsequent inflationary era.

\section{Mode matching as a cure for IR divergences}\label{tre}


\noindent We now assume that the solution to the mode equation is characterized by two cosmological phases determined by two different values of the index $\mu$, which we denote by $\mu$ and $\alpha$. The global evolution of $Q$ then is
\bea
\hspace{-2mm}Q_\mu \! \!&=& \! a_\mu^{-1}(t)\,u_{\mu}(z)\ , \,\,\,t \le t_i \ ,\label{e:ini}   \\
\hspace{-2mm}Q_\alpha\! \! &=& \! a_\alpha^{-1}(t)\left[E_\alpha(k)u_\alpha(z)+F_\alpha(k)u_\alpha^*(z)\right] ,\,\,   t_i<t.
\label{e:ini2}
\eea
As we mentioned above, 
Eq.(\ref{e:ini}) is a sensible vacuum initial condition only for a  
radiation-dominated or inflationary phase.
We shall be interested in the first case later on, but we keep our discussion general in the
beginning.

We impose that $a_{\mu}(t_{i})=a_{\alpha}(t_{i})$ and $\dot a_{\mu}(t_{i})=\dot a_{\alpha}(t_{i})$. This can be realized in the following way. Consider the equation $\epsilon=-\dot H/H^{2}=$ const. The general solution for the Hubble parameter is
\bea
H=\frac{H_0}{1 +\epsilon H_0 (t-t_0)},
\eea
which implies that
\bea
a(t)=a_0 \left[1+\epsilon H_0 (t-t_0)\right]^{1/\epsilon}\ .
\eea
In these expressions,  $H_{0}$ and $a_{0}$ denote the arbitrary values of $H(t)$ and $a(t)$ at $t=t_0$. 
Now, suppose that
\bea
a(t)=\left\{\begin{array}{c} a_{1}\left[1+\epsilon_1 H_{1}(t-t_1)\right]^{1/\epsilon_{1}}  \quad t\leq t_{i}\\\\   
a_{2}\left[1+\epsilon_2 H_{2}(t-t_2)\right]^{1/\epsilon_{2}}\quad t>t_{i}\end{array}\right.\ .
\eea
If we consider the case for which $H_1=H_2$ the match can be easily imposed setting 
$t_1=t_2=t_i$ and $a_1=a_2$. On the contrary, if $H_1 \neq H_2$, by
 imposing continuity of $H(t)$ across $t_{i}$ we find 
\bea
t_{i}={H_{2}-H_{1}\over H_{1}H_{2}(\epsilon_{2}-\epsilon_{1})}\ ,
\eea
together with the consistency condition
\bea
\epsilon_1 t_1=\epsilon_2 t_2=c\ ,
\eea
where $c$ is an arbitrary constant.
The continuity of $a(t)$ across $t_{i}$ further implies that 
\bea
\frac{a_2}{a_1}=\left[\frac{H_2 \epsilon_2-H_1 \epsilon_1}{\epsilon_2-\epsilon_1}
-c H_1 H_2\right]^{1/\epsilon_1-1/\epsilon_2}
\frac{H_1^{1/\epsilon_2}}
{H_2^{1/\epsilon_1}}\ .
\eea
We are interested in a transition between a radiation-dominated Universe ($\epsilon_{1}=2$) and a slow-roll inflationary one ($0<\epsilon_{2}\ll1$). In this case, the scale factor has the form 
$a(t)\sim t^{1/2}$ before the match and $a(t)\simeq a_0 e^{H_0(t-t_0)+\frac{\dot{H}_0}{2}(t-t_0)^2}$ 
after the match. So considering an inflationary phase that starts at the time of the match the scale 
factor takes the following form 
\bea
a(t)=\left\{\begin{array}{c} (t/t_i)^{1/2}  \quad t\leq t_{i}\\\\   
e^{H_i(t-t_i)+\frac{\dot{H}_i}{2}(t-t_i)^2} \quad t>t_{i}\end{array}\right.\,,
\eea
with $H_i=\frac{1}{2 t_i}$. The inflationary expression for the scale factor is exact for a quadratic potential, $V = m^2\phi^2/2$, and is a first order expansion in the slow-roll parameters for other choices of the inflationary potential. For the numerical results shown below we always choose a quadratic potential.

We now determine the coefficients $E_{\alpha}$ and $F_{\alpha}$ in Eq.~(\ref{e:ini2}), by solving the linear system
\bea\label{linsys}
\left\{\begin{array}{c}Q_\mu(t_i)=Q_\alpha(t_i)  \\\\ \dot Q_\mu(t_i)=\dot Q_\alpha(t_i)\end{array}\right. \ ,
\eea
with the help of the result
\bea\label{identity}
\dot z(t) \simeq -{k\over a(t)}\ ,
\eea
which holds whenever $\epsilon$ is nearly constant, and using the 
Wronskian identity ($\gamma=\alpha,\mu$)
\bea
u_\gamma(t)\dot u_\gamma^*(t)-u_\gamma^*(t)\dot u_\gamma(t)={i\over a_\gamma(t)} \ ,
\eea
we find
\bea\label{EaFa}
\left.\begin{array}{c} E_\alpha=ia(t_i)\left(u_\alpha^*\dot u_\mu-\dot u_\alpha^* u_\mu\right)_{t=t_i}\ , \\\\ F_\alpha=ia(t_i)\left(\dot u_\alpha u_\mu-u_\alpha\dot u_\mu\right)_{t=t_i} \,,\end{array}\right.
\eea
and it is easy to check that $|E_\alpha|^2-|F_\alpha|^2=1$.

In the rest of the paper, we assume that the state of the Universe preceding the inflation is  dominated by a scalar field with an exponential  potential leading to a radiation-like expansion. To realize this, it is sufficient to take \cite{Marozzi}
\bea\non
\phi(t)&=&M_{\rm pl} \ln \left(t\over t_{i}\right) +\phi_{i}\,,\\
V&=&{M_{\rm pl}^2 \over 4 t_i^2}\exp\left[-\frac{2(\phi(t)-\phi_{i})}{M_{\rm pl}}\right]\ ,
\eea
where $\phi_{1}$ is an arbitrary constant. In fact, one can check that $\epsilon=2$, $\eta =4$ thus, from Eq.\ (\ref{Eq_mukhanov_2}) one obtains
\bea
\label{radsol}
Q={i\over a\sqrt{2k}}e^{-iz}\,.
\eea
Another advantage of this point of view is that the Mukhanov formalism is well-defined for all times, and there is no need to adjust Eq.\ (\ref{Eq_mukhanov_2}) with the speed of sound.

The infrared divergences become apparent when one calculates the two-point function $G$ at coincident points
\bea
G\propto\int dkk^{2}|Q_\alpha|^2 \, .
\eea
In the small $z$ limit, one finds that
\bea\label{smallz}
u_\alpha(z)\simeq \sqrt{\pi z\over 4 k}{\Gamma(\alpha)\over i\pi}\left(2\over z\right)^{\alpha}\simeq -u_\alpha^*(z) \ ,\\\non\\
\dot u_\alpha(z)+\dot u_\alpha^*(z)\simeq {\dot z\over \alpha \Gamma(\alpha)2^{\alpha}}\sqrt{\pi\over k}\left({1\over 2}+\alpha\right)z^{\alpha-{1\over 2}}\ ,
\eea
thus $|Q_\alpha|^2\propto |u_\alpha|^2 |E_\alpha-F_\alpha|^2 \propto k^{-2\alpha}|E_\alpha-F_\alpha|^2$. By using Eq.\ (\ref{smallz}), its time derivative and similar expressions for $u_\mu$, 
one finds that, for $z(t_i)\ll 1$, $|E_\alpha-F_\alpha|^2\sim k^{2(\alpha-\mu)}$ and
\bea
G\propto \int dk k^{2(1-\mu)}\ .
\eea
This result first shows that the low-$k$ behavior of the integral is independent of $\alpha$, namely of whether the Universe accelerates or decelerates after the transition at $t=t_i$.   Then, it is clear that the integral converges provided $\mu<3/2$, which excludes configurations where the Universe accelerates for $t<t_i$, such as de Sitter or slow-roll. On the contrary, if the Universe begins in a radiation-dominated phase $\mu=1/2$ the IR convergence is guaranteed. These calculations confirm
the statement that IR divergences cannot, in general, develop during a smooth evolution 
of the Universe \cite{ford}. Similar results were found in  \cite{proko2}, where the evolution of a test scalar field was studied during a smooth transition from a decelerating and expanding Universe to an inflationary one.


\section{UV impact of the mode matching }\label{quattro}


\noindent The above results show that mode matching can solve the IR problem if the very first phase of the Universe is of non-inflationary type. The infrared end of the spectrum is sensible only to this phase, and cannot be changed by the future evolution (provided $\epsilon$ is nearly constant). On the ultraviolet side, we know that adiabatic subtraction of appropriate terms can cancel the UV divergences. The main problem, which has not been addressed yet, is the evolution of these counter-terms through the transition at $t=t_{i}$.
We will address this issue in the next section, but first let us see if the divergent structure after the match is influenced by the presence of the match itself.

After the transition, the wave function is no longer in a Bunch-Davies state, as it is represented by the function 
\bea\label{gensol}
Q_\alpha=a_\alpha^{-1}(t)\left[E_\alpha(k)u_\alpha(z)+F_\alpha(k)u_\alpha^*(z)\right] \ ,
\eea
where $E_\alpha$ and $F_\alpha$ are given by Eq.\ (\ref{EaFa}). We wish to evaluate $|Q_\alpha|^2$ in the ultraviolet regime. For large $z$ we have \cite{abramo}
\bea\label{largez}
H^{(1)}_\mu=\sqrt{2\over \pi z}\,e^{i\chi_\mu}\,\left[S_\mu+iR_\mu\right]\ ,
\eea
where
\bea\label{PQ}\non
R_\mu&=&1-{(4\mu^2-1)(4\mu^2-9)\over 128z^2}+\ldots\ ,\\ 
S_\mu&=&{4\mu^2-1\over 8z}+\ldots\ ,\\
\chi_\mu&=&z-\pi\left({\mu\over 2}+{1\over 4}\right)\ ,
\eea
and similar expressions for $H^{(2)}_\alpha$.
With these expansions and Eq.\ (\ref{EaFa}), calling $z_\mu$ the value of $z$ for $t\leq t_i$ and 
$z_\alpha$ the value for $t>t_i$, we find that
\bea\label{phiadiab}
&&\!\!\!\!|Q_{\alpha}|^{2}\simeq{1\over 2ka^{2}}\Bigg\{  1+{4\alpha^{2}-1\over 8z_{\alpha}^{2}}+\\
&&\!\!\!-\left[\frac{4\alpha^{2}-1}{\bar z_{\alpha}^{2}}- \frac{4\mu^{2}-1}{\bar z_{\mu}^{2}}\right] 
{\cos (2z_{\alpha}-2\bar z_{\alpha})\over 8} \Bigg\},\non
\eea
where the barred  quantities are those evaluated at $t=t_{i}$. The third term in square brackets only appears because of the match at $t=t_{i}$, and consistently vanishes when  $\mu=\alpha$. This term is not divergent in the UV for $t > t_{i}$.
Before the match, the ultraviolet structure of the modes is
\bea
|Q_{\mu}|^{2}&\simeq&{1\over 2ka^{2}}\Bigg[ 1+{(4\mu^{2}-1)\over 8z_{\mu}^{2}}\Bigg]\ ,
\eea
and shows the usual logarithmic and quadratic divergences (except for the radiation-dominated case considered, where 
$|\mu|=1/2$ and the logarithmic one is not present). From Eq.\ (\ref{phiadiab}) we see that the divergent structure after the match is not altered by the presence of the match at $t=t_{i}$. 
As a check of the method, one easily verifies that at $t=t_{i}$ the ultraviolet limit of $|Q_{\alpha}|^{2}$ coincides with the one of $|Q_{\mu}|^{2}$, as required by the first condition in Eqs.\ (\ref{linsys}).


\section{Adiabatic expansion through the match}\label{cinque}


\noindent The UV regularization of the two-point function can be achieved by subtracting appropriate counter-terms from the divergent integral. These quantities can be constructed by solving, through an 
adiabatic expansion at the appropriate order, the mode equation \cite{BD,ParkerAE}. In order to be consistent with the mode matching,  we wish compute the counter-terms and study their evolution through the phase transition at $t=t_{i}$. 
Let us write the adiabatic solution, before and after the match,  as
\bea
Q^{\rm ad}_{\mu}(t)&=& {1 \over a_\mu\sqrt{2W_{k \mu}}}
\exp\left(i\int_{t_i}^t {dt\over a_\mu}W_{k\mu}\right) \nonumber \\ 
Q^{\rm ad}_{\alpha}(t)&=&{E_{\alpha}^{\rm ad}\over a_\alpha\sqrt{2W_{k \alpha}}}\exp\left(i\int_{t_i}^t {dt\over a_\alpha}W_{k \alpha}\right)+ \nonumber \\
&+&{F_{\alpha}^{\rm ad}\over a_\alpha \sqrt{2W_{k \alpha}}}\exp\left(-i\int_{t_i}^t {dt\over a_\alpha}
W_{k \alpha}\right)\ ,
\eea 
where the superscript ``ad'' stands for adiabatic, and where $W_{k s}$ must satisfy the equation
\bea
W_{k \gamma}^2=\Omega_{k \gamma}^{2}-{1\over 2}\left({W_{k \gamma}''\over W_{k \gamma}}-{W_{k \gamma}'^{2}\over W_{k \gamma}^{2}}\right)\ ,
\eea
where $\gamma=\mu/\alpha$. 
The generalized frequency $\Omega_{k \gamma}$ is defined as
\be
\Omega_{k \gamma}^{2}=k^{2}+a_\gamma^{2} M_{\gamma}^2
\ee
with
\be
M_{\gamma}^2=V_{\phi\phi}
-{1\over 6}\left[R-6\left( {2a'''\over a^{2}a'}-4{a'^{2}\over a^{4}}-{2a''^{2}\over a^{2}a'^{2}}  \right)\right]\,,
\ee
where the underscript $\gamma$ is neglected in the r.h.s..
For our purposes, it is sufficient to keep the terms up to the second adiabatic order (namely, with two time derivatives). We stress that also the term $V_{\phi\phi}$, should be considered of second order as well 
\footnote{We wish to remark here that the  effective mass $V_{\phi\phi}$ was already treated as a second order adiabatic quantity in \cite{us}. 
This work was the first one to state explicitly that $V_{\phi\phi}$ is a second order 
adiabatic term.
We disagree with what is written in \cite{Agullo}: in fact, the expressions of the adiabatic counter-terms for tensor and scalar gauge invariant perturbations described in \cite{Agullo} were already given in \cite{us}.
For example, a little notational change shows that the adiabatic counter-term represented in Eq.\ (3.33) of \cite{Agullo}  is exactly equivalent to Eq.\ (43) of \cite{us}. Similarly, the expression for the subtraction from the non-renormalized part of the adiabatic one, for scalar perturbations, given in Eq.\ (4.4) of \cite{Agullo} coincides with Eq.\ (45) of \cite{us}.}.

By solving again the system (\ref{linsys}) for the adiabatic case, we find the coefficients
\bea
E_{\alpha}^{\rm ad}\!&=&-{(\bar\Omega_{k \alpha}+\bar \Omega_{k \mu})\over 2 (\bar \Omega_{k \alpha}\bar \Omega_{k \mu})^{1/2}}\, ,\\
F_{\alpha}^{\rm ad}\!&=&-{(\bar\Omega_{k \alpha}-\bar \Omega_{k \mu})\over 2 (\bar \Omega_{k \alpha}\bar \Omega_{k \mu})^{1/2}}\,,
\eea
where only terms up to the second adiabatic order have been retained. 
With these, we find
\bea\non
|Q^{\rm ad}_{\alpha}|^{2}&=&{1\over 4a_\alpha^{2}\Omega_{k \alpha}}\Bigg[\left({\bar\Omega_{k \alpha}\over \bar\Omega_{k \mu}}+{\bar\Omega_{k \mu}\over \bar\Omega_{k \alpha}}\right)+\left({\bar\Omega_{k \alpha}\over \bar\Omega_{k \mu}}-{\bar\Omega_{k \mu}\over \bar\Omega_{k \alpha}}\right)\times\\
&\times&\cos \left(2\int_{t_{i}}^{t}{dt'\over a_\alpha(t')}\Omega_{k \alpha}\right)\Bigg]\,.
\eea
Expanding up to the second adiabatic order this expression can be written as
\bea\non\label{WKBexp}
|Q^{\rm ad}_{\alpha}|^{2}&=& \frac{1}{2 a_\alpha^2 k}  \left(1-\frac{1}{2}
\frac{a_\alpha^2 M_{\alpha}^2}{k^2}\right) \left[1+\frac{1}{2} \frac{a_i^2}{k^2}
\left(\bar{M}_{\alpha}^2-\bar{M}_{\mu}^2\right)\right. \nonumber \\
& & \left. \cos \left(2\int_{t_{i}}^{t}{dt'\over a_\alpha(t')}\Omega_{k \alpha}\right)\right]\ .
\eea
The first part of this expansion cancel the divergent terms in Eq.\ (\ref{phiadiab}) in the ultraviolet limit. So, using such expression we can have a consistent regularization of the field fluctuations 
through the time of the matching and beyond.

\section{Observational signatures}\label{sei}

\noindent We want now to evaluate possible signatures of the match considered 
on the spectrum of the scalar perturbations.
Every mode exiting the horizon at the time $t_{\rm ex}$ satisfies the relation $a(t_{\rm ex})H(t_{\rm ex})=k$. In typical inflationary models, modes that exit  about 60 e-folds before the end inflation correspond to scales that are observable today. We wish to compute the spectrum associated with these modes and verify whether the pre-match phase can have left some signature. Thus, we calculate the coefficients $E_{\alpha}$ and $F_{\alpha}$ at the time $t=t_{i}$ according to Eqs.\ (\ref{EaFa}) and $|Q_{\alpha}|^{2}$ according to Eq.\ (\ref{gensol}).
Then we can compute the scalar power spectrum defined by Eq.(\ref{spectrum}). 

Let us consider two limiting cases, the one where $t_i=t_{\rm ex}$ and, as consequences, 
$a(t_i) H(t_i)=a(t_{\rm ex})H(t_{\rm ex})=k$ and the one where $t_i\ll t_{\rm ex}$ with 
$a(t_i) H(t_i)\ll a(t_{\rm ex})H(t_{\rm ex})=k$. In this second case $\bar{z}_\mu,\bar{z}_\alpha \gg 1$ and, using 
Eqs.\ (\ref{largez},\ref{PQ}), one finds
\be
|E_\alpha|^2=1+{\cal O}\left(\frac{1}{\bar{z}_\alpha^3}\right)\,\,\,,\,\,\,
|F_\alpha|^2={\cal O}\left(\frac{1}{\bar{z}_\alpha^3}\right)\,,
\ee
\be
E_\alpha F_\alpha^*={\cal O}\left(\frac{1}{\bar{z}_\alpha^2}\right)\,.
\ee
At the leading order, we have
\begin{eqnarray}
|Q_{\alpha}|^{2}&=&\frac{1}{a^2}\left[\left(|E_\alpha|^2 + |F_\alpha|^2\right) |u_\alpha|^2 +
E_\alpha F_\alpha^* u_\alpha^2+ E_\alpha^* F_\alpha u_\alpha^{*\,2}\right]
\nonumber \\
&\simeq& \frac{1}{a^2} |u_\alpha|^2\,,
\end{eqnarray}
so we recover the standard slow-roll behavior, and we have no observational consequences of the match at the leading order.

On the contrary, in the case  $t_i=t_{\rm ex}$, the contribution of $E_\alpha$ and $F_\alpha$ give sensible corrections to the power spectrum. To see this, let us first evaluate the spectrum exactly at the horizon exit.  In the standard case we have 
\bea
|Q_\alpha^{\rm st}|^2= {\pi z_{\alpha}\over 4ka^{2}}|H_{\alpha}^{(1)}(z_{\alpha})|^{2}\,,
\eea 
with $z_{\alpha}(k=a H) \simeq (1+\epsilon)$. 
By expanding with respect to the slow-roll parameters, the leading term 
is given by the Hankel function with $\alpha=3/2$, so 
\bea
|H_{\alpha}^{(1)}(z_{\alpha})|^{2}\sim {2\over \pi z_{\alpha}}\left(1+{1\over z_{\alpha}^{2}}\right)\,,
\eea
and
\bea
P_{\zeta}^{\,\rm st}(k=a H)\simeq{1\over M_{\rm pl}^2\epsilon }\left(H\over 2\pi\right)^{2}\,.
\eea
On the other hand, if one considers the matching condition $Q_\alpha(t_i)=Q_\mu(t_i)$ together with
\be
Q_\mu=\frac{1}{a} \sqrt{\frac{\pi}{4 a H}} H_{\mu}^{(2)}\left(\frac{k}{a H}\right)\ ,
\ee
with $\mu=1/2$ (radiation domination),
the modified spectrum at the horizon exit becomes
\bea
P_{\zeta}(k=a H)={1\over 2 M_{\rm pl}^2\epsilon }\left(H\over 2\pi\right)^{2}\,,
\eea
so it is reduced  by about $50\%$.

Let us now instead consider the case in which the spectrum is evaluated several e-folds (already $5$ e-folds are sufficient) after the horizon exit, namely when $k \ll a H$. 
This means that the argument of the Hankel function is relatively small and we can make the approximation  $|H_{\alpha}^{(1)}(z_{\alpha})|^{2}\sim 2/(\pi z_{\alpha}^3)$. 
Then, the spectrum has the well-known value
\bea
P_{\zeta}^{\,\rm st}(k\ll a H)={1\over 2M_{\rm pl}^2\epsilon }\left(H \over 2\pi\right)^{2}\,.
\eea
As before, the introduction of the matching conditions together with the requirement that the fluctuations exit the horizon at the time of the match ($k=a_i H_i$) changes our result. In the limit   $k \ll a H$ we have that $u_\alpha^*\simeq -u_\alpha$ and, as in Sec.\ III, one obtains
\be
|Q_\alpha|^2\simeq \frac{|u_\alpha|^2}{a^2} |E_\alpha-F_\alpha|^2 \,.
\label{SpectrumMod2}
\ee
The first part on the r.h.s. gives the standard contribution to the spectrum, 
while $|E_\alpha-F_\alpha|^2$ calculated on $k=a_i H_i$ is
 a numerical coefficient, which is independent of the initial condition and nearly equal to $0.38$. 
So the spectrum, in the presence of the match, is reduced by about the $62\%$.

\begin{figure}[ht]
\centering
\includegraphics[width=80mm,height=70mm]{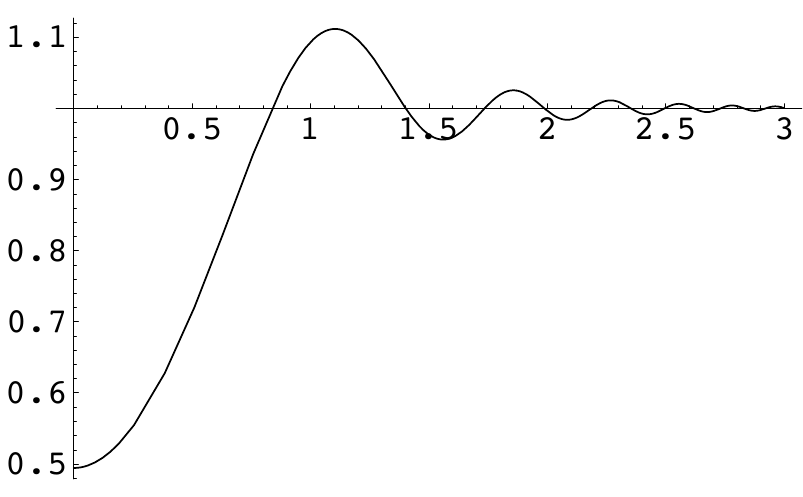}
\caption{\label{fig2}
$P_\zeta(k=a H)/P_\zeta^{\,\rm st}(k=a h)$ is shown for a $m^2 \phi^2$ potential versus the number 
of e-folds $N$ between the beginning of the inflation and the time where the fluctuation considered crosses the horizon. The spectra are calculated at the time of horizon exit.  }
\end{figure}

\begin{figure}[ht]
\centering
\includegraphics[width=80mm,height=80mm]{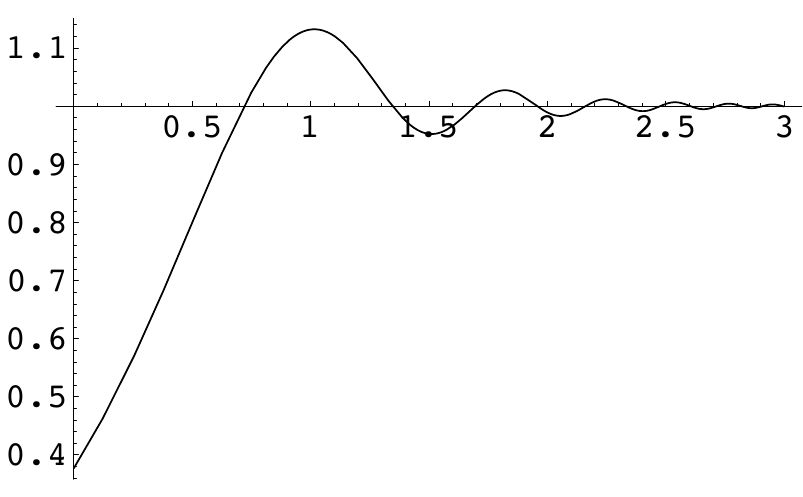}
\caption{\label{fig3}
$P_\zeta(k\ll a H)/P_\zeta^{\,\rm st}(k\ll a H)$ is shown for a $m^2 \phi^2$ potential
versus the number of e-folds $N$ between the beginning of the inflation and the time where the fluctuation considered crosses the horizon. The spectra are calculated several e-folds after the horizon exit.}
\end{figure}

The general behavior of the modified spectrum for these two cases is shown 
in Figs.\ \ref{fig2} and \ref{fig3}. Here, we plot the ratio of the modified power spectrum with 
respect to the standard one for the case when the inflationary potential is
$V=\frac{1}{2} m^2 \phi^2$, with the match described in section III and an initial condition fixed by a value of  $H(t_i)$, which guarantees near $60$  e-folds of inflation.  

\begin{figure}[ht]
\centering
\includegraphics[width=80mm,height=70mm]{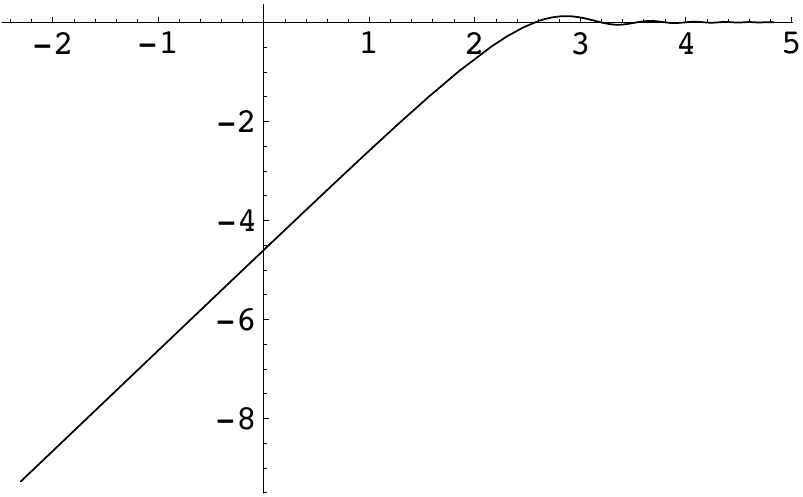}
\caption{\label{fig4}
$\log \left(P_\zeta(k\ll a H)/P_\zeta^{\,\rm st}(k\ll a H)\right)$ versus the $\log(k/m)$ is shown for a $m^2 \phi^2$ model with nearly $60$ e-folds of inflation. The range of $k$ goes from $0.1 m$ to the value that exits the 
horizon $3$ e-folds after the beginning of the inflation (for $k=a(t_i) H(t_i)$ we have
$\log(k/m)\simeq 1.85$). The spectra are calculated several e-folds after the beginning of inflation.}
\end{figure}

To conclude this section, let us consider fluctuations with wave numbers 
$k<a(t_i) H(t_i)$, namely fluctuations that were outside the horizon at the beginning of inflation, and still are today. Although their spectrum is not directly accessible to observations, these modes can have an 
effect on non-Gaussianities produced at second order, which will be the subject of a future work. If we consider the limit $k\ll a(t_i) H(t_i)\leq a(t) H(t)$ we have the same condition that leads to Eq.\ (\ref{SpectrumMod2}) but now, as shown in Sec.\  III, $|E_\alpha-F_\alpha|^2\sim k^{2(\alpha-\mu)}$ and the spectrum changes from nearly scale-invariant to very blue, 
$P_\zeta(k)\sim k^2$.
In Fig.~\ref{fig4} we show the general behavior of this spectrum, 
calculated several e-folds after the exit of the 
observable fluctuations, going 
from $k\ll a(t_i) H(t_i)$ to $k\gg a(t_i) H(t_i)$.
As discussed in \cite{us}, the renormalization of the spectrum by adiabatic subtraction at the horizon exit seems meaningless (see also next section). Therefore, we omit the evaluation at horizon exit of the adiabatic counter-terms.




\section{Validity range of the adiabatic subtraction}\label{sette}


\noindent In this section, we would like to study the validity range of the adiabatic expansion. 
The adiabatic counter-terms obtained with an adiabatic expansion, as the one shown in 
Sec.\ \ref{cinque}, can be found also with a de Witt-Schwinger (dWS) point-spitting. However, we shall see that the dWS series is no longer valid in the slow-roll approximation as one approaches horizon exit. From the point of view of the adiabatic subtraction, this is equivalent to the loss of adiabaticity.

To show this, let us briefly recall the dWS procedure (for more details, see \cite{BD,BunchParker}). The starting point is the local expansion of the metric in Riemann Normal Coordinates (RNC), which can be regarded as a constructive proof of the local flatness theorem \cite{Poisson,BD}. On a smooth manifold, we can always expand the metric around a given point $P$ as
\bea
g_{\mu\nu}(Q)=\eta_{\mu\nu}+{1\over 3}R_{\mu\alpha\nu\beta}(P)y^{\alpha}y^{\beta}+\cdots\ ,
\eea     
where $y^{\alpha}$ are the RNC with origin at $P$, $\eta_{\mu\nu}$ is the Minkowski metric, and the dots represent higher curvature terms. If the vector with components $p^{\alpha}$ is the tangent at $P$ to the geodesics that joins the points $P$ and $Q$, we can define the RNC as $y^{\alpha}=\tau p^{\alpha}$, where $\tau$ is an affine parameter along the geodesics. It is clear, from the construction itself, that the validity of the RNC patch is restricted to a region where geodesics do not intersect. There is however another constraint on the typical size $L$ of the volume around $P$ where RNC are valid, namely that the curvature is not rapidly changing. These validity regimes can be stated more precisely as  \cite{Nesterov}
\bea\label{bounds}
L\ll {\rm min}\left({1\over |R_{\mu\alpha\nu\beta}(P)|^{1/2}}, { |R_{\mu\alpha\nu\beta}(P)|\over |R_{\mu\alpha\nu\beta,\gamma}(P)|}\right) \ ,
\eea
where the first bound comes from the size of volume without intersections and the second from the size of volume where the curvature is slowly varying. If these conditions are met, one can define a local Fourier transform operator and write  the two-point function as \cite{BunchParker}
\bea
G(P,Q)=\int {d^4k\over (2\pi)^4}\,e^{iky}\tilde G(k)\ .
\eea 
For a minimally coupled massless scalar field, the function $\tilde G(k)$ can be expanded as
\bea\label{dsexp}
\tilde G(k)={1\over k^2}\left[1+{a^{2}R\over 6k^2}+\cdots\right]\ ,
\eea
with the expansion valid only when $|a^{2}R/k^2|\ll 1$. On a spatially flat FLRW background, this condition reads
\bea
{a^2H^2\over k^2}\left[2+{\dot H\over H^2}\right]\ll 1\,.
\eea
In the particular case of slow-roll inflation we see that the validity of the RNC expansion is set by the constraint $aH\ll  k$, so it is not correct to calculate spectra at the horizon exit, $aH= k$, with this renormalization method. 

We can verify this finding by looking directly at the bounds imposed by the inequalities (\ref{bounds}). For a FLRW background with metric (\ref{metric}) one has 
\bea
R_{0i0j}= (\epsilon-1)a^2H^2\delta_{ij}\ , \quad R_{ijij}=a^4H^2\ ,
\eea
where latin indices label spatial coordinates and there is no summation over repeated indices. It follows that, in the slow-roll regime, when $\dot\epsilon\simeq 0$  and $|\epsilon|\ll 1$, one has
\bea
\Big|{R_{0i0j}\over \dot R_{0i0j}}\Big|\simeq {1\over |1-\epsilon|H}\ ,\quad \Big|{R_{ijij}\over \dot R_{ijij}}\Big|\simeq {1\over 2|2-\epsilon|H}\,,
\eea
and this  implies that  the dWS expansion is valid in a region of size $L\ll 1/(aH)$, namely much smaller than the comoving Hubble radius that decreases during inflation. To enlarge the validity domain, one should consider more terms in the expansion. But this would spoil the rule that the number of adiabatic counter-terms has to be two, in order to just cure the quadratic and the logarithmic divergence. 

These considerations reinforce the findings of \cite{us}, namely that adiabatic or dWS subtraction are not suitable to renormalize the two-point function, and hence the power spectrum, at and after the horizon exit.

\section{Conclusions}\label{otto}

\noindent In this paper we have examined the simultaneous regularization of both infrared and ultraviolet divergences of  cosmological perturbations, treated as quantum fields on a curved background. The infrared regularization is performed by matching slow roll inflation to a radiation dominated,  pre-inflationary phase. 
We do not take this particular phase too serious, but our procedure shows, with this concrete example, that a previous phase with a well defined initial vacuum can well regularize inflationary infrared divergences. It will be interesting to study in the future, whether higher order perturbations, which are relevant e.g. for non-Gaussianities, will depend on the details of this infrared regularization.
Furthermore, since our modification happens before the onset of inflation and just removes
power in the modes which never enter the horizon during inflation, we expect it
to  regularize the infrared also for inflationary models  which deviate from slow roll as, for example, warm inflation \cite{warm}. The details of the matching and the possible observational signatures might however be somewhat modified.  

To regularize the ultraviolet divergence we employ the usual subtraction of adiabatic counter-terms. 
The first result is that the mode matching does not introduce new ultraviolet divergences. In addition, 
the adiabatic expansion is well defined through the match despite what one might fear because of the discontinuity of the Ricci scalar $R$. In fact, in our model $R$ is not continuous through the transition as it contains second derivatives of the scale factor. Therefore, one might expect that the de Witt-Schwinger expansion (\ref{dsexp}), which is equivalent to the adiabatic expansion, be no longer well defined. 
On the contrary, we find that this is not the case.

 Finally, we argue that the adiabatic expansion is 
not valid at and after the horizon exit, by looking more carefully at the construction of the counter-terms by the 
de Witt-Schwinger point-splitting method. This reinforces our opinion that ultraviolet regularization cannot leave 
any observable imprint. However, it seems possible that the mode matching in the infrared
has left some observational trace, but only if it occurs very close to horizon exit of the scales of interest.
\vspace{0.5cm}

\noindent
{\bf Acknowledgement:}
G.M. wishes to thank Fabio Finelli for comments on the manuscript.
This work is supported by the Swiss National Science Foundation.


\end{document}